# An Achievable Rate Region for a Two-Relay Network with Receiver-Transmitter Feedback

Mohammad Ali Tebbi[1], *Student Member, IEEE,* Mahtab Mirmohseni[2], Mahmoud Ahmadian Attari[1], and Mohammad Reza Aref[2]

[1] Coding and Cryptography Lab. (CCL), Electrical and Computer Engineering Department, K.N. Toosi University of Tech., Tehran, Iran
[2] Information Systems and Security Lab. (ISSL), Electrical Engineering Department, Sharif University of Tech., Tehran, Iran
E-mail: m.a.tebbi@ee.kntu.ac.ir, mirmohseni@ee.sharif.edu, mahmoud@eetd.kntu.ac.ir, aref@sharif.edu

*Abstract*—**We consider a relay network with two relays and a feedback link from the receiver to the sender. To obtain the achievability result, we use compress-and-forward and random binning techniques combined with deterministic binning and restricted decoding. Moreover, we use joint decoding technique to decode the relays' compressed information to achieve a higher rate in the receiver.**

*Keywords- relay network; receiver-transmitter feedback; compress-and-forward; achievable rate.*

## I. INTRODUCTION

The relay channel with feedback was first considered by Cover and El Gamal in [1]. In their channel model, there were feedback links from the receiver to both the sender and the relay and from the relay to the sender, referred to as complete feedback [2]. It was shown that complete feedback or partial feedback from the receiver to the relay makes the relay channel a physically degraded relay channel, thus the cut-set upper bound is achievable [1]. The relay channel with partial feedback from the receiver or the relay to the sender has been investigated in [3] and [4]. It has been shown that neither the receiver-transmitter nor the relay-transmitter feedback can improve the capacity of the physically degraded and the semi-deterministic relay channel [4].

The relay network was first introduced in [5], where the capacity of a general relay network with complete feedback, i.e. feedback from the receiver to the all relays and the sender, and from each relay to the sender and the previous relays, has been derived and has been shown that the complete feedback can increase the capacity. Partial feedback from the receiver to the relays and from each relay to the previous ones make the relay network a physically degraded relay network, thus cannot increase the capacity [6].

In [7] and [8] relay networks with parallel relaying have been considered. In parallel relaying, there is no straight link between the sender and the receiver. Also, the relays do not interchange any information. In [9], some cooperative strategies for relay networks have been discussed and reviewed. Additionally, the authors in [9] generalize the compress-and-forward strategy to relay networks.

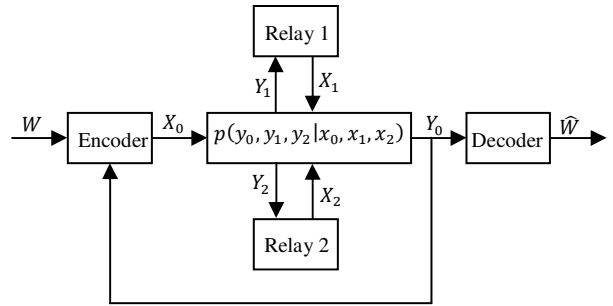

Fig. 1. Two-Relay Network with Receiver-Transmitter Feedback

Symmetric relay network has been introduced in [10]. In a symmetric two-relay network, there is a straight link from the sender to the receiver and each relay can completely decode the message transmitted by the other relay in addition to the part of the message transmitted by the sender. However, no achievable rate expressions have yet been obtained for relay networks with partial feedback to the sender.

In this paper we consider a relay network with two relays and partial feedback from the receiver to the sender. In our proposed model depicted in Fig. 1, there is a feed-forward link from the sender to the receiver and a feedback link from the receiver to the sender. Both of the relays help the receiver to solve his uncertainty about the sender. Each relay tries to send the information about the messages to the receiver as much as possible through the direct link between the relay and the receiver. However, the relays have no information interchange. The feedback link develops some cooperation between the sender and the relays, since the sender knows the relays' compressed information transmitted to the receiver. For this model, we use compress-and-forward coding scheme [1] and random partitioning [11] combined with deterministic partitioning and restricted decoding [12] to obtain an achievable rate.

The rest of the paper is organized as fallows. Section II, introduces the network model and definitions. In Section III, we present the achievable rate obtained for the model. In Section IV, the achievability of the rate reported in Section III is proved, and In Section V, we conclude the paper.

## II. PRELIMINARIES AND DEFINITIONS

In this paper, upper case letters (e.g., $X$) are used to denote Random Variables (RVs) while their realizations are denoted by lower case letters (e.g., $x$). The alphabet of a random variable $X$ will be designated by a calligraphic letter $\mathcal{X}$. $x_i^{j(k)}$ indicates the sequence of $\left(x_{i,1}^{(k)}, x_{i,2}^{(k)}, \ldots, x_{i,j}^{(k)}\right)$, where $k$ denotes the block number of transmission. $p_X(x)$ denotes the probability mass function of $X$ on a set $\mathcal{X}$, where occasionally subscript $X$ is omitted.

*Definition 1:* The discrete memoryless two-relay network with receiver-transmitter feedback ($\mathcal{X}_0 \times \mathcal{X}_1 \times \mathcal{X}_2$, $p(y_0, y_1, y_2|x_0, x_1, x_2)$, $\mathcal{Y}_0 \times \mathcal{Y}_1 \times \mathcal{Y}_2$) depicted in Fig. 1, consists of a sender $X_0 \in \mathcal{X}_0$, a receiver $Y_0 \in \mathcal{Y}_0$, relay senders $X_1 \in \mathcal{X}_1$ and $X_2 \in \mathcal{X}_2$, relay receivers $Y_1 \in \mathcal{Y}_1$ and $Y_2 \in \mathcal{Y}_2$, and a family of conditional probability mass functions $p(y_0, y_1, y_2|x_0, x_1, x_2)$ on $\mathcal{Y}_0 \times \mathcal{Y}_1 \times \mathcal{Y}_2$ one for each $(x_0, x_1, x_2) \in \mathcal{X}_0 \times \mathcal{X}_1 \times \mathcal{X}_2$. An $(M, n)$ code for this network consists of a message set $\mathcal{M} = \{1, 2, \ldots, M\}$, an encoding function $x_0: \mathcal{M} \times \mathcal{Y}_0^{t-1} \to \mathcal{X}_0$ for $t = 1, \ldots, n$, a set of relay functions $\{f_{ij}\}$ such that $x_{ij} = f_{ij}(y_{i1}, y_{i2}, \ldots, y_{i,j-1})$ for $1 \leq j \leq n$, $1 \leq i \leq 2$, and a decoding function $g: \mathcal{Y}_0 \to \mathcal{M}$. A rate $R = \frac{1}{n} \log M$ is achievable if there is an $(M, n)$ code with $M \geq 2^{nR}$ such that $\bar{p}_e^n = \Pr\{g(Y_0^n) \neq W | W = w, w \in \mathcal{M}\} < \varepsilon$, for any $\varepsilon > 0$ and for all $n$ sufficiently large.

## III. MAIN RESULTS

In this section we present an achievable rate concerning a relay network with two relays and a feedback link from receiver to sender depicted in Fig. 1.

*Theorem 1:* Consider the discrete memoryless relay network with two relays ($\mathcal{X}_0 \times \mathcal{X}_1 \times \mathcal{X}_2$, $p(y_0, y_1, y_2|x_0, x_1, x_2)$, $\mathcal{Y}_0 \times \mathcal{Y}_1 \times \mathcal{Y}_2$) and causal noiseless feedback from the receiver to the sender defined in section II. Then, the rate $R$ defined by

$$R = \sup_{p(v_1, v_2, x_0, x_1, x_2, y_1, y_2, \hat{y}_1, \hat{y}_2, y_0)} \{I(X_0; Y_0, \hat{Y}_1, \hat{Y}_2, X_1, X_2|V_1, V_2) + I(V_1, X_1; V_2, X_2)\} \quad (1)$$

is achievable subject to the constraints

$$I(\hat{Y}_1; Y_1|V_1, X_1) < \min \{I(X_1; X_0, Y_0, V_2, X_2|V_1) + I(V_1; V_2, X_2) + I(\hat{Y}_1; X_0, Y_0|V_1, X_1),$$
$$I(V_1; Y_0, V_2) + I(X_1; Y_0, V_2, X_2|V_1) + I(V_1; V_2, X_2) + I(\hat{Y}_1; Y_0|V_1, X_1)\} \quad (2)$$

$$I(\hat{Y}_2; Y_2|V_2, X_2) < \min \{I(X_2; X_0, Y_0, V_1, X_1|V_2) + I(V_2; V_1, X_1) + I(\hat{Y}_2; X_0, Y_0|V_2, X_2),$$
$$I(V_2; Y_0, V_1) + I(X_2; Y_0, V_1, X_1|V_2) + I(V_2; V_1, X_1) + I(\hat{Y}_2; Y_0|V_2, X_2)\} \quad (3)$$

$$I(\hat{Y}_1; Y_1|V_1, X_1) + I(\hat{Y}_2; Y_2|V_2, X_2) <$$
$$\min \{I(X_1, X_2; X_0, Y_0|V_1, V_2) + I(V_1; V_2)$$
$$+ I(\hat{Y}_1; X_0, Y_0|V_1, X_1) + I(\hat{Y}_2; X_0, Y_0|V_2, X_2),$$
$$I(V_1; Y_0, V_2) + I(V_2; Y_0, V_1) + I(X_1, X_2; Y_0|V_1, V_2)$$
$$+ I(V_1; V_2) + I(\hat{Y}_1; Y_0|V_1, X_1) + I(\hat{Y}_2; Y_0|V_2, X_2),$$
$$I(V_1, V_2; Y_0) + I(X_1, X_2; Y_0|V_1, V_2) + I(V_1; V_2)$$
$$+ I(\hat{Y}_1; Y_0|V_1, X_1) + I(\hat{Y}_2; Y_0|V_2, X_2),$$
$$I(V_1, V_2; Y_0) + I(X_1; Y_0, V_2, X_2|V_1) + I(V_1; V_2, X_2)$$
$$+ I(X_2; Y_0, V_1, X_1|V_2) + I(V_2; V_1, X_1) +$$
$$I(\hat{Y}_1; Y_0|V_1, X_1) + I(\hat{Y}_2; Y_0|V_2, X_2)\} \quad (4)$$

Where the supremum is taken over all joint p.m.f on $\mathcal{V}_1 \times \mathcal{V}_2 \times \mathcal{X}_0 \times \mathcal{X}_1 \times \mathcal{X}_2 \times \mathcal{Y}_1 \times \mathcal{Y}_2 \times \hat{\mathcal{Y}}_1 \times \hat{\mathcal{Y}}_2 \times \mathcal{Y}_0$ of the form

$p(v_1, v_2, x_0, x_1, x_2, y_1, y_2, \hat{y}_1, \hat{y}_2, y_0)$
$= p(v_1)p(v_2)p(x_1|v_1)p(x_2|v_2)p(x_0|v_1, v_2)$
$\cdot p(y_0, y_1, y_2|x_0, x_1, x_2)p(\hat{y}_1|y_1, x_1, v_1)p(\hat{y}_2|y_2, x_2, v_2)$.

*Remark 1:* The auxiliary random variables $V_1$ and $V_2$ construct a two level cooperation between the sender and the two relays.

*Remark 2:* The first terms in the constraints (2), (3), and (4) are dictated by the decoding procedure at the sender and reflect the minimal compression ratio sustainable by the sender as it attempts to decode the compressed data sent by the first, second, and both of relays to the receiver, respectively.

*Remark 3:* The second terms in the constraints (2), (3), and the second to the forth terms in the constraint (4) are dictated by the decoding procedure at the receiver and reflect the minimal compression ratio sustainable by the receiver taking into account the help it gets from both the sender and the first relay, both the sender and the second relay, and the sender, the first and the second relays together, respectively.

*Remark 4:* In Appendix C we use joint technique to decode the relays' compressed information. Using this technique does not increase the achievable rate expression (1) but can increase the minimal compression ratios sustainable by the sender and the receiver. Thus, it can make some relaxations on the constraints (2)-(4) and can improve the achievable rate.

## IV. PROOFS

To prove of the achievability of rate (1), we use compress-and-forward coding scheme [1] based on block Markov superposition encoding and random binning proof of the source coding theorem of Slepian-Wolf [11]. In addition, to decrease complexity of list coding techniques we utilize the restricted decoding and nonrandom binning, instead of list decoding and lexicographical indexing, introduced in [12] for multiple-access channel with partial feedback. In all decoding steps, we use joint decoding technique, except when the sender and the receiver decode the compressed information of the relays, in order to avoid the complexity and make the results more understandable.

*Proof of theorem 1:* Consider a block Markov encoding scheme where a sequence of $B-1$ messages $w^{(b)} \in [1, 2^{nR}]$ for $b = 1, 2, \ldots, B-1$ is transmitted in $B+1$ blocks, each of $n$ symbols. As $B \to \infty$, the rate $R(B-1)/(B+1)$ is arbitrarily close to $R$.

*Random coding:* Generate $2^{n(R_{011}+R_{012})}$ i.i.d sequences $v_1^n$, each with probability $p(v_1^n) = \prod_{i=1}^n p(v_{1i})$ and label them $v_1^n(w_{01})$ where $w_{01} = (w_{011}, w_{012})$, $w_{011} \in [1, 2^{nR_{011}}]$, and $w_{012} \in [1, 2^{nR_{012}}]$. Generate $2^{n(R_{021}+R_{022})}$ i.i.d sequences $v_2^n$, each with probability $p(v_2^n) = \prod_{i=1}^n p(v_{2i})$ and label them $v_2^n(w_{02})$ where $w_{02} = (w_{021}, w_{022})$, $w_{021} \in [1, 2^{nR_{021}}]$, and $w_{022} \in [1, 2^{nR_{022}}]$. For each $v_1^n$, generate $2^{nR_{s_1}}$ i.i.d sequences $x_1^n$, each with probability $p(x_1^n | v_1^n) = \prod_{i=1}^n p(x_{1i}|v_{1i})$ and label them $x_1^n(s_1, w_{01})$ where $s_1 \in [1, 2^{nR_{s_1}}]$. For each $v_2^n$, generate $2^{nR_{s_2}}$ i.i.d sequences $x_2^n$, each with probability $p(x_2^n | v_2^n) = \prod_{i=1}^n p(x_{2i}|v_{2i})$ and label them $x_2^n(s_2, w_{02})$ where $s_2 \in [1, 2^{nR_{s_2}}]$. For each $(v_1^n, v_2^n)$, generate $2^{nR}$ i.i.d sequences $x_0^n$, each with probability $p(x_0^n|v_1^n, v_2^n) = \prod_{i=1}^n p(x_{0i}|v_{1i}, v_{2i})$ and label them $x_0^n(w, w_{01}, w_{02})$, where $w \in [1, 2^{nR}]$. For each $x_1^n$, choose $2^{n\hat{R}_1}$ i.i.d sequences $\hat{y}_1^n$, each with probability $p(\hat{y}_1^n|x_1^n, v_1^n) = \prod_{i=1}^n p(\hat{y}_{1i}|x_{1i}, v_{1i})$ where, for $x_1 \in \mathcal{X}_1$, $\hat{y}_1 \in \hat{\mathcal{Y}}_1$ and $v_1 \in \mathcal{V}_1$ we define

$$p(\hat{y}_1^n|x_1^n, v_1^n) = \sum_{x_0, y_0, y_1, y_2} p(x_0|v_1, v_2) p(y_0, y_1, y_2|x_0, x_1, x_2) p(\hat{y}_1|y_1, x_1, v_1)$$

Label them $\hat{y}_1^n(z_1|s_1, w_{01})$ where $z_1 \in [1, 2^{n\hat{R}_1}]$. For each $x_2^n$, choose $2^{n\hat{R}_2}$ i.i.d sequences $\hat{y}_2^n$, each with probability $p(\hat{y}_2^n|x_2^n, v_2^n) = \prod_{i=1}^n p(\hat{y}_{2i}|x_{2i}, v_{2i})$ where, for $x_2 \in \mathcal{X}_2$, $\hat{y}_2 \in \hat{\mathcal{Y}}_2$ and $v_2 \in \mathcal{V}_2$ we define

$$p(\hat{y}_2^n|x_2^n, v_2^n) = \sum_{x_0, y_0, y_1, y_2} p(x_0|v_1, v_2) p(y_0, y_1, y_2|x_0, x_1, x_2) p(\hat{y}_2|y_2, x_2, v_2)$$

Label them $\hat{y}_2^n(z_2|s_2, w_{02})$ where $z_2 \in [1, 2^{n\hat{R}_2}]$.

*Partitioning:*
1. Randomly partition the set $\{1, 2, \ldots, 2^{n\hat{R}_1}\}$ into $2^{nR_{s_1}}$ cells $\mathcal{S}_{s_1}$ for $s_1 \in \{1, 2, \ldots, 2^{nR_{s_1}}\}$, and the set $\{1, 2, \ldots, 2^{n\hat{R}_2}\}$ into $2^{nR_{s_2}}$ cells $\mathcal{S}_{s_2}$ for $s_2 \in \{1, 2, \ldots, 2^{nR_{s_2}}\}$.
2. Randomly partition each cell of size $2^{n(\hat{R}_1 - R_{s_1})}$ into $2^{nR_{012}}$ subcells $\mathcal{S}_{w_{012}}$ for $w_{012} \in \{1, 2, \ldots, 2^{nR_{012}}\}$, and each cell of size $2^{n(\hat{R}_2 - R_{s_2})}$ into $2^{nR_{022}}$ subcells $\mathcal{S}_{w_{022}}$ for $w_{022} \in \{1, 2, \ldots, 2^{nR_{022}}\}$.
3. Create a partition over the set $\{1, 2, \ldots, 2^{nR_{s_1}}\}$ with $2^{nR_{011}}$ disjoint cells $\mathcal{S}_{w_{011}}$ for $w_{011} \in \{1, 2, \ldots, 2^{nR_{011}}\}$, each containing $2^{n(R_{s_1} - R_{011})}$ elements, and a partition over the set $\{1, 2, \ldots, 2^{nR_{s_2}}\}$ with $2^{nR_{021}}$ disjoint cells $\mathcal{S}_{w_{021}}$ for $w_{021} \in \{1, 2, \ldots, 2^{nR_{021}}\}$, each containing $2^{n(R_{s_2} - R_{021})}$ elements.

*Remark 5:* The third partition referred to as deterministic partition and we will use it for restricted decoding [12].

In joint decoding procedure, the decoder decodes a pair of bin numbers which their contents are jointly $\epsilon$-typical.

*Encoding:*
Let $w^{(b)}$ be the new message to be sent in block $b$. Assume that $\left(\hat{y}_1^n\left(z_1^{(b-1)}\middle|s_1^{(b-1)}, w_{01}^{(b-1)}\right), y_1^{n(b-1)}, x_1^{n(b-1)}, v_1^{n(b-1)}\right)$ are jointly $\epsilon$-typical, and $\left(\hat{y}_2^n\left(z_2^{(b-1)}\middle|s_2^{(b-1)}, w_{02}^{(b-1)}\right), y_2^{n(b-1)}, x_2^{n(b-1)}, v_2^{n(b-1)}\right)$ are jointly $\epsilon$-typical. Then, the codewords transmitted by the first and the second relays in block $b$ are

$$x_1^n\left(s_1^{(b)}, w_{01}^{(b)}\right) = x_1^n\left(s_1\left(z_1^{(b-1)}\right), \left(w_{011}\left(s_1^{(b-1)}\right), w_{012}\left(z_1^{(b-2)}\right)\right)\right)$$

and

$$x_2^n\left(s_2^{(b)}, w_{02}^{(b)}\right) = x_2^n\left(s_2\left(z_2^{(b-1)}\right), \left(w_{021}\left(s_2^{(b-1)}\right), w_{022}\left(z_2^{(b-2)}\right)\right)\right),$$

respectively, while the codeword transmitted by the sender is

$$x_0^n\left(w^{(b)}, \hat{w}_{01E}^{(b)}, \hat{w}_{02E}^{(b)}\right) = x_0^n\Big(w^{(b)}, \left(w_{011}\left(\hat{s}_{1E}^{(b-1)}\right), w_{012}\left(\hat{z}_{1E}^{(b-2)}\right)\right),$$
$$\left(w_{021}\left(\hat{s}_{2E}^{(b-1)}\right), w_{022}\left(\hat{z}_{2E}^{(b-2)}\right)\right)\Big).$$

*Decoding:*
*At the sender:* At the end of block $b$ for $b = 1, 2, \ldots, B-1$, the sender finds a unique pair $\left(\hat{s}_{1E}^{(b)}, \hat{s}_{2E}^{(b)}\right)$ such that $\left(v_1^n\left(\hat{w}_{01E}^{(b)}\right), v_2^n\left(\hat{w}_{02E}^{(b)}\right), x_1^n\left(\hat{s}_{1E}^{(b)}, \hat{w}_{01E}^{(b)}\right), x_2^n\left(\hat{s}_{2E}^{(b)}, \hat{w}_{02E}^{(b)}\right), x_0^n\left(w^{(b)}, \hat{w}_{01E}^{(b)}, \hat{w}_{02E}^{(b)}\right), y_0^{n(b)}\right)$ are jointly $\epsilon$-typical. This determines $w_{011}\left(\hat{s}_{1E}^{(b)}\right)$ and $w_{021}\left(\hat{s}_{2E}^{(b)}\right)$, which are transmitted by the sender in block $b+1$. For sufficiently large $n$, the decoding error in this step is arbitrarily small if

$$R_{s_1} < I(X_1; X_0, Y_0, V_2, X_2|V_1) + I(V_1; V_2, X_2) \quad (5)$$
$$R_{s_2} < I(X_2; X_0, Y_0, V_1, X_1|V_2) + I(V_2; V_1, X_1) \quad (6)$$
$$R_{s_1} + R_{s_2} < I(X_1, X_2; X_0, Y_0|V_1, V_2) + I(V_1; V_2) \quad (7)$$

For the proof of (5)-(7) see Appendix A.
Then, the sender calculates his ambiguity sets $\mathcal{L}_{1E}\left(y_0^{n(b-1)}\right)$ and $\mathcal{L}_{2E}\left(y_0^{n(b-1)}\right)$ of $\hat{z}_{1E}^{(b-1)}$ and $\hat{z}_{2E}^{(b-1)}$ such that $\hat{z}_{1E}^{(b-1)} \in \mathcal{L}_{1E}\left(y_0^{n(b-1)}\right)$ and $\hat{z}_{2E}^{(b-1)} \in \mathcal{L}_{2E}\left(y_0^{n(b-1)}\right)$ if $\left(v_1^n\left(\hat{w}_{01E}^{(b-1)}\right), x_1^n\left(\hat{s}_{1E}^{(b-1)}, \hat{w}_{01E}^{(b-1)}\right), y_0^{n(b-1)}, x_0^n\left(w^{(b-1)}, \hat{w}_{01E}^{(b-1)}, \hat{w}_{02E}^{(b-1)}\right), \hat{y}_1^n\left(z_{1E}^{(b-1)}\middle|\hat{s}_{1E}^{(b-1)}, \hat{w}_{01E}^{(b-1)}\right)\right)$ and $\left(v_2^n\left(\hat{w}_{02E}^{(b-1)}\right), x_2^n\left(\hat{s}_{2E}^{(b-1)}, \hat{w}_{02E}^{(b-1)}\right), y_0^{n(b-1)}, x_0^n\left(w^{(b-1)}, \hat{w}_{01E}^{(b-1)}, \hat{w}_{02E}^{(b-1)}\right), \hat{y}_2^n\left(z_{2E}^{(b-1)}\middle|\hat{s}_{2E}^{(b-1)}, \hat{w}_{02E}^{(b-1)}\right)\right)$ are jointly $\epsilon$-typical, respectively. Next, he declares that $\hat{z}_{1E}^{(b-1)}$ and $\hat{z}_{2E}^{(b-1)}$ were sent in block $b-1$ iff there are unique

$$\hat{z}_{1E}^{(b-1)} \in \mathcal{S}_{\hat{s}_{1E}^{(b)}} \cap \mathcal{L}_{1E}\left(y_0^{n(b-1)}\right)$$

and

$$\hat{z}_{2E}^{(b-1)} \in \mathcal{S}_{\hat{s}_{2E}^{(b)}} \cap \mathcal{L}_{2E}\left(y_0^{n(b-1)}\right),$$

respectively. For sufficiently large $n$, the decoding error in this step is arbitrarily small if

$$\hat{R}_1 < I(\hat{Y}_1; X_0, Y_0|V_1, X_1) + R_{s_1} \quad (8)$$
$$\hat{R}_2 < I(\hat{Y}_2; X_0, Y_0|V_2, X_2) + R_{s_2} \quad (9)$$

For the proof of (8) and (9) see Appendix B.

The above two decoding steps in the sender provide a two level cooperation between the sender and both of the relays.

*At the receiver:* At the end of block $b+1$, the receiver looks for a unique pair $\left(\widehat{w}_{01D}^{(b+1)}, \widehat{w}_{02D}^{(b+1)}\right)$ such that $\left(v_1^n\left(\widehat{w}_{01D}^{(b+1)}\right), v_2^n\left(\widehat{w}_{02D}^{(b+1)}\right), y_0^{n(b+1)}\right)$ are jointly $\epsilon$-typical. For sufficiently large $n$, the decoding error in this step is arbitrarily small if

$$R_{011} + R_{012} < I(V_1; Y_0, V_2) \qquad (10)$$
$$R_{021} + R_{022} < I(V_2; Y_0, V_1) \qquad (11)$$
$$R_{011} + R_{012} + R_{021} + R_{022} < I(V_1, V_2; Y_0) \qquad (12)$$

The proof of (10)-(12) is similar to those for (5)-(7) in Appendix A.

Then, the receiver considers block $b$ and chooses a unique pair $\left(\hat{s}_{1D}^{(b)}, \hat{s}_{2D}^{(b)}\right)$ such that $\left(v_1^n\left(\widehat{w}_{01D}^{(b)}\right), v_2^n\left(\widehat{w}_{02D}^{(b)}\right), x_1^n\left(\hat{s}_{1D}^{(b)}, \widehat{w}_{01D}^{(b)}\right), x_2^n\left(\hat{s}_{2D}^{(b)}, \widehat{w}_{02D}^{(b)}\right), y_0^{n(b)}\right)$ are jointly $\epsilon$-typical. For sufficiently large $n$, the decoding error in this step is arbitrarily small if

$$R_{s_1} < I(X_1; Y_0, V_2, X_2|V_1) + I(V_1; V_2, X_2) + R_{011} \qquad (13)$$
$$R_{s_2} < I(X_2; Y_0, V_1, X_1|V_2) + I(V_2; V_1, X_1) + R_{021} \qquad (14)$$
$$R_{s_1} + R_{s_2} < I(X_1, X_2; Y_0|V_1, V_2) + I(V_1; V_2) + R_{011} + R_{021} \qquad (15)$$

The proof of (13)-(15) is similar to those for (5)-(7) in Appendix A. Here $\widehat{w}_{01D}^{(b)}$ and $\widehat{w}_{02D}^{(b)}$ were already determined in decoding steps similar to those for (10)-(12), and the decoding is restricted to $\left(\hat{s}_{1D}^{(b)}, \hat{s}_{2D}^{(b)}\right)$ inside cells $\left(\widehat{w}_{011}^{(b+1)}, \widehat{w}_{021}^{(b+1)}\right)$. This step is similar to the restricted decoding in [12].

Then, the receiver considers block $b-1$ and calculates his ambiguity sets $\mathcal{L}_{1D}\left(y_0^{n(b-1)}\right)$ and $\mathcal{L}_{2D}\left(y_0^{n(b-1)}\right)$ of $\hat{z}_{1D}^{(b-1)}$ and $\hat{z}_{2D}^{(b-1)}$ such that $\hat{z}_{1D}^{(b-1)} \in \mathcal{L}_{1D}\left(y_0^{n(b-1)}\right)$ and $\hat{z}_{2D}^{(b-1)} \in \mathcal{L}_{2D}\left(y_0^{n(b-1)}\right)$ if $\left(v_1^n\left(\widehat{w}_{01D}^{(b-1)}\right), x_1^n\left(\hat{s}_{1D}^{(b-1)}, \widehat{w}_{01D}^{(b-1)}\right), y_0^{n(b-1)}, \hat{y}_1^n\left(z_{1D}^{(b-1)}|\hat{s}_{1D}^{(b-1)}, \widehat{w}_{01D}^{(b-1)}\right)\right)$ and $\left(v_2^n\left(\widehat{w}_{02D}^{(b-1)}\right), x_2^n\left(\hat{s}_{2D}^{(b-1)}, \widehat{w}_{02D}^{(b-1)}\right), y_0^{n(b-1)}, \hat{y}_2^n\left(z_{2D}^{(b-1)}|\hat{s}_{2D}^{(b-1)}, \widehat{w}_{02D}^{(b-1)}\right)\right)$ are jointly $\epsilon$-typical, respectively. The receiver declares that $\hat{z}_{1D}^{(b-1)}$ and $\hat{z}_{2D}^{(b-1)}$ were sent in block $b-1$ iff there are unique

$$\hat{z}_{1D}^{(b-1)} \in \mathcal{S}_{\hat{s}_{1D}^{(b)}} \cap \mathcal{S}_{\widehat{w}_{012}^{(b+1)}} \cap \mathcal{L}_{1D}\left(y_0^{n(b-1)}\right)$$

and

$$\hat{z}_{2D}^{(b-1)} \in \mathcal{S}_{\hat{s}_{2D}^{(b)}} \cap \mathcal{S}_{\widehat{w}_{022}^{(b+1)}} \cap \mathcal{L}_{2D}\left(y_0^{n(b-1)}\right),$$

respectively. For sufficiently large $n$, the decoding error in this step is arbitrarily small if

$$\hat{R}_1 < I(\hat{Y}_1; Y_0|V_1, X_1) + R_{s_1} + R_{012} \qquad (16)$$
$$\hat{R}_2 < I(\hat{Y}_2; Y_0|V_2, X_2) + R_{s_2} + R_{022} \qquad (17)$$

The proof of (16) and (17) is similar to those for (9) and (10) in Appendix B.

Then, the receiver declares that $\widehat{w}^{(b-1)}$ was sent in block $b-1$ if $\left(v_1^n\left(\widehat{w}_{01D}^{(b-1)}\right), v_2^n\left(\widehat{w}_{02D}^{(b-1)}\right), x_1^n\left(\hat{s}_{1D}^{(b-1)}, \widehat{w}_{01D}^{(b-1)}\right), x_2^n\left(\hat{s}_{2D}^{(b-1)}, \widehat{w}_{02D}^{(b-1)}\right), x_0^n\left(\widehat{w}^{(b-1)}, \widehat{w}_{01D}^{(b-1)}, \widehat{w}_{02D}^{(b-1)}\right), y_0^{n(b-1)}, \hat{y}_1^n\left(\hat{z}_{1D}^{(b-1)}|\hat{s}_{1D}^{(b-1)}, \widehat{w}_{01D}^{(b-1)}\right), \hat{y}_2^n\left(\hat{z}_{2D}^{(b-1)}|\hat{s}_{2D}^{(b-1)}, \widehat{w}_{02D}^{(b-1)}\right)\right)$ are jointly $\epsilon$-typical. Using packing lemma [13], for sufficiently large $n$, the decoding error in this step is arbitrarily small if

$$R < I(X_0; Y_0, \hat{Y}_1, \hat{Y}_2, X_1, X_2|V_1, V_2) + I(V_1, X_1; V_2, X_2)$$

which provides the achievable rate (1).

*At the relays:* The first relay upon receiving $y_1^{n(b)}$ decides that $z_1^{(b)}$ is received if $\left(\hat{y}_1^n\left(z_1^{(b)}|s_1^{(b)}, w_{01}^{(b)}\right), y_1^{n(b)}, v_1^n\left(w_{01}^{(b)}\right), x_1^n\left(s_1^{(b)}, w_{01}^{(b)}\right)\right)$ are jointly $\epsilon$-typical. For sufficiently large $n$, the decoding error in this step is arbitrarily small if

$$\hat{R}_1 > I(\hat{Y}_1; Y_1|V_1, X_1) \qquad (18)$$

The decoding step at the second relay is similar to those at the first relay. Thus, we have

$$\hat{R}_2 > I(\hat{Y}_2; Y_2|V_2, X_2) \qquad (19)$$

For the proof of (18) and (19), see [14].

Combining (5)-(19) and applying Fourier-Motzkin elimination, constraints (2)-(4) are derived.

## V. CONCLUSION

In this paper, we considered a relay network with two relays and partial feedback configuration from the receiver to the transmitter. We used compress-and-forward coding scheme and random binning combined with deterministic binning and restricted decoding to obtain an achievable rate. We applied individual and joint techniques to decode the relays' compressed information and showed that joint decoding gives some relaxation on the rate constraints.

## APPENDIX A

## PROOF OF (5)-(7)

Define the event $E_{s_E}$ in block $b+1$ as $\left(v_1^n\left(w_{01}^{(b+1)}\right), v_2^n\left(w_{02}^{(b+1)}\right), x_0^n\left(w^{(b+1)}, w_{01}^{(b+1)}, w_{02}^{(b+1)}\right), x_1^n\left(s_1^{(b+1)}, w_{01}^{(b+1)}\right), x_2^n\left(s_2^{(b+1)}, w_{02}^{(b+1)}\right), y_0^{n(b+1)}\right)$ are jointly $\epsilon$-typical. By the union bound, the probability of error at the sender in block $b+1$ for the event $E_{s_E}$ can be upper bounded as

$$p_{E_{s_E}}^e = \Pr\left(E_{s_E}^c \cup \bigcup_{(\tilde{s}_{1E}, \tilde{s}_{2E}) \neq (s_1^{(b+1)}, s_2^{(b+1)})} E_{s_E}\right) \leq$$
$$\Pr(E_{s_E}^c) + \sum_{(\tilde{s}_{1E}, \tilde{s}_{2E}) \neq (s_1^{(b+1)}, s_2^{(b+1)})} \Pr(E_{s_E}) \qquad (20)$$

Based on Asymptotic Equipartition Property (AEP) [14], the first term in (20) tends to zero as $n \to \infty$, where $E_{s_E}^c$ denotes the complement of the event $E_{s_E}$. For the second term in (20), there are three cases which by packing lemma [13], we have

$$\Pr\left\{E_{s_E}|\tilde{s}_{1E} \neq s_1^{(b+1)}\right\} \leq 2^{-n(I(X_1; X_0, Y_0, V_2, X_2|V_1) + I(V_1; V_2, X_2) - 6\epsilon)} \qquad (21)$$

$$\Pr\{E_{S_E}|\tilde{s}_{2E} \neq s_2^{(b+1)}\} \leq 2^{-n(I(X_2;X_0,Y_0,V_1,X_1|V_2)+I(V_2;V_1,X_1)-6\epsilon)} \quad (22)$$
$$\Pr\{E_{S_E}|\tilde{s}_{1E} \neq s_1^{(b+1)}, \tilde{s}_{2E} \neq s_2^{(b+1)}\} \leq$$
$$2^{-n(I(X_1,X_2;X_0,Y_0|V_1,V_2)+I(V_1;V_2)-6\epsilon)} \quad (23)$$

Considering (20) and (21)-(23), $p^e_{E_{S_E}}$ goes to zero as $n \to \infty$ if (5)-(7) hold. ∎

## APPENDIX B

### PROOF OF (8) AND (9)

Define the event $E_{z_{1E}}$ in block b+1 as
$$z_1^{(b)} \in \mathcal{S}_{s_1^{(b+1)}} \cap \mathcal{L}_{1E}(y_0^{n(b)})$$

We assume that the decoding in the previous block was successful (the event $F^{(b)^c}$) and define $\psi(z_{1E}|y_0^{n(b)}) = 1$ if $(v_1^n(w_{01}^{(b)}), x_0^n(w^{(b)}, w_{01}^{(b)}, w_{02}^{(b)}), y_0^{n(b)}, x_1^n(s_1^{(b)}, w_{01}^{(b)}), \hat{y}_1^n(z_{1E}|s_1^{(b)}, w_{01}^{(b)}))$ are jointly $\epsilon$-typical. Otherwise, define $\psi(z_{1E}|y_0^{n(b)}) = 0$.

The cardinality of $\mathcal{L}_{1E}(y_0^{n(b)})$ is the random variable $\|\mathcal{L}_{1E}(y_0^{n(b)})\| = \sum_{z_{1E}} \psi(z_{1E}|y_0^{n(b)})$ and

$$\mathbb{E}\{\|\mathcal{L}_{1E}(y_0^{n(b)})\| \, |F^{(b)^c}\} = \mathbb{E}\{\psi(z_1^{(b)}|y_0^{n(b)})|F^{(b)^c}\} + \sum_{z_{1E} \neq z_1^{(b)}} \mathbb{E}\{\psi(z_{1E}|y_0^{n(b)})|F^{(b)^c}\},$$

where $\mathbb{E}$ denotes the expectation operator. By packing lemma [13], we have
$$\mathbb{E}\{\psi(z_{1E}|y_0^{n(b)})|F^{(b)^c}\} \leq 2^{-n(I(\hat{Y}_1;X_0,Y_0|V_1,X_1)-6\epsilon)}$$

Then,
$$\mathbb{E}\{\|\mathcal{L}_{1E}(y_0^{n(b)})\| \, |F^{(b)^c}\} \leq$$
$$1 + (2^{n\hat{R}_1} - 1)(2^{-n(I(\hat{Y}_1;X_0,Y_0|V_1,X_1)-6\epsilon)}) \quad (24)$$

The conditioning on $F^{(b)^c}$ implies that $z_1^{(b)} \in \mathcal{L}_{1E}(y_0^{n(b)})$, and the assumption of occurring the event $E_{S_E}$ implies that $\hat{s}_{1E}^{(b+1)} = s_1^{(b+1)}$ which means $z_1^{(b)} \in \mathcal{S}_{s_1^{(b+1)}}$. thus, the probability of error for the event $E_{z_{1E}}$ is
$$p^e_{E_{z_{1E}}} \leq \mathbb{E}\{\sum_{z_{1E} \neq z_1^{(b)}} \Pr(z_{1E} \in \mathcal{S}_{s_1^{(b+1)}}|F^{(b)^c})\}$$
$$\leq \mathbb{E}\{\|\mathcal{L}_{1E}(y_0^{n(b)})\| \, |F^{(b)^c}\} 2^{-nR_{s_1}}. \quad (25)$$

From (24) and (25), and for sufficiently large $n$, (8) is proved. The proof of (9) is similar to those for (8).

## APPENDIX C

### JOINT DECODING THE COMPRESSED INFORMATION

For joint decoding the relays' compressed information in the sender, after each block $b$ for $b = 1,2,\ldots,B$, the sender calculates his ambiguity set $\mathcal{L}_E(y_0^{n(b-1)})$ of $(\hat{z}_{1E}^{(b-1)}, \hat{z}_{2E}^{(b-1)})$ such that $(v_1^n(\hat{w}_{01E}^{(b-1)}), v_2^n(\hat{w}_{02E}^{(b-1)}), x_1^n(\hat{s}_{1E}^{(b-1)}, \hat{w}_{01E}^{(b-1)}), x_2^n(\hat{s}_{2E}^{(b-1)}, \hat{w}_{02E}^{(b-1)}), y_0^{n(b-1)}, x_0^n(w^{(b-1)}, \hat{w}_{01E}^{(b-1)}, \hat{w}_{02E}^{(b-1)}), \hat{y}_1^n(z_{1E}^{(b-1)}|\hat{s}_{1E}^{(b-1)}, \hat{w}_{01E}^{(b-1)}), \hat{y}_2^n(z_{2E}^{(b-1)}|\hat{s}_{2E}^{(b-1)}, \hat{w}_{02E}^{(b-1)}))$ are jointly $\epsilon$-typical. The sender declares that $(\hat{z}_{1E}^{(b-1)}, \hat{z}_{2E}^{(b-1)})$ were sent in block $b-1$ iff there is a unique pair
$$(\hat{z}_{1E}^{(b-1)}, \hat{z}_{2E}^{(b-1)}) \in \mathcal{S}_{(\hat{s}_{1E}^{(b)}, \hat{s}_{2E}^{(b)})} \cap \mathcal{L}_E(y_0^{n(b-1)})$$

For sufficiently large $n$, the decoding error in this step is arbitrarily small if

$$\hat{R}_1 < I(\hat{Y}_1;X_0,Y_0,V_2,X_2,\hat{Y}_2|V_1,X_1) + I(V_1,X_1;V_2,X_2) + R_{s_1} \quad (26)$$
$$\hat{R}_2 < I(\hat{Y}_2;X_0,Y_0,V_1,X_1,\hat{Y}_1|V_2,X_2) + I(V_1,X_1;V_2,X_2) + R_{s_1} \quad (27)$$
$$\hat{R}_{11} + \hat{R}_2 < I(\hat{Y}_1,\hat{Y}_2;X_0,Y_0|V_1,V_2,X_1,X_2) + I(V_1,X_1;V_2,X_2) + R_{s_1} \quad (28)$$
$$\hat{R}_1 < I(\hat{Y}_1;X_0,Y_0,V_2,X_2,\hat{Y}_2|V_1,X_1) + I(V_1,X_1;V_2,X_2) + R_{s_2} \quad (29)$$
$$\hat{R}_2 < I(\hat{Y}_2;X_0,Y_0,V_1,X_1,\hat{Y}_1|V_2,X_2) + I(V_1,X_1;V_2,X_2) + R_{s_2} \quad (30)$$
$$\hat{R}_{11} + \hat{R}_2 < I(\hat{Y}_1,\hat{Y}_2;X_0,Y_0|V_1,V_2,X_1,X_2) + I(V_1,X_1;V_2,X_2) + R_{s_2} \quad (31)$$

For the proof of (26)-(31), define the event $E_{z_E}$ in block $b+1$ as:
$$(z_1^{(b)}, z_2^{(b)}) \in \mathcal{S}_{(s_1^{(b+1)}, s_2^{(b+1)})} \cap \mathcal{L}_E(y_0^{n(b)})$$

We assume that the decoding in the previous block was successful (the event $F^{(b)^c}$) and define $\psi((z_{1E}, z_{2E})|y_0^{n(b)}) = 1$ if $(v_1^n(w_{01}^{(b)}), v_2^n(w_{02}^{(b)}), x_0^n(w^{(b)}, w_{01}^{(b)}, w_{02}^{(b)}), y_0^{n(b)}, x_1^n(s_1^{(b)}, w_{01}^{(b)}), x_2^n(s_2^{(b)}, w_{02}^{(b)}), \hat{y}_1^n(z_{1E}|s_1^{(b)}, w_{01}^{(b)}), \hat{y}_2^n(z_{2E}|s_2^{(b)}, w_{02}^{(b)}))$ are jointly $\epsilon$-typical. Otherwise, define $\psi((z_{1E}, z_{2E})|y_0^{n(b)}) = 0$.

The cardinality of $\mathcal{L}_E(y_0^{n(b)})$ is the random variable $\|\mathcal{L}_E(y_0^{n(b)})\| = \sum_{(z_{1E},z_{2E})} \psi((z_{1E}, z_{2E})|y_0^{n(b)})$ and

$$\mathbb{E}\{\|\mathcal{L}_E(y_0^{n(b)})\| \, |F^{(b)^c}\} = \mathbb{E}\{\psi((z_1^{(b)}, z_2^{(b)})|y_0^{n(b)})|F^{(b)^c}\} + \sum_{(z_{1E},z_{2E}) \neq (z_1^{(b)}, z_2^{(b)})} \mathbb{E}\{\psi((z_{1E}, z_{2E})|y_0^{n(b)})|F^{(b)^c}\},$$

where $\mathbb{E}$ denotes the expectation operator. For the second term in (32), there are three cases, which by packing lemma [13], we have
$$\mathbb{E}\{\psi((z_{1E} \neq z_1^{(b)}, z_{2E})|y_0^{n(b)})|F^{(b)^c}\}$$
$$\leq 2^{-n(I(\hat{Y}_1;X_0,Y_0,V_2,X_2,\hat{Y}_2|V_1,X_1)+I(V_1,X_1;V_2,X_2)-6\epsilon)}$$
$$\mathbb{E}\{\psi((z_{1E}, z_{2E} \neq z_2^{(b)})|y_0^{n(b)})|F^{(b)^c}\}$$
$$\leq 2^{-n(I(\hat{Y}_2;X_0,Y_0,V_1,X_1,\hat{Y}_1|V_2,X_2)+I(V_1,X_1;V_2,X_2)-6\epsilon)}$$
$$\mathbb{E}\{\psi((z_{1E} \neq z_1^{(b)}, z_{2E} \neq z_2^{(b)})|y_0^{n(b)})|F^{(b)^c}\}$$
$$\leq 2^{-n(I(\hat{Y}_1,\hat{Y}_2;X_0,Y_0|V_1,V_2,X_1,X_2)+I(V_1,X_1;V_2,X_2)-6\epsilon)}$$

Then,
$$\mathbb{E}\{\|\mathcal{L}_E(y_0^{n(b)})\| \, |F^{(b)^c}\}$$
$$\leq 1 + (2^{n\hat{R}_1} - 1)(2^{-n(I(\hat{Y}_1;X_0,Y_0,V_2,X_2,\hat{Y}_2|V_1,X_1)+I(V_1,X_1;V_2,X_2)-6\epsilon)})$$
$$+ (2^{n\hat{R}_2} - 1)(2^{-n(I(\hat{Y}_2;X_0,Y_0,V_1,X_1,\hat{Y}_1|V_2,X_2)+I(V_1,X_1;V_2,X_2)-6\epsilon)})$$
$$+ (2^{n(\hat{R}_1+\hat{R}_2)} - 1)(2^{-n(I(\hat{Y}_1,\hat{Y}_2;X_0,Y_0|V_1,V_2,X_1,X_2)+I(V_1,X_1;V_2,X_2)-6\epsilon)}). \quad (32)$$

The conditioning on $F^{(b)^c}$ implies that $(z_1^{(b)}, z_2^{(b)}) \in \mathcal{L}_E(y_0^{n(b)})$, and the assumption of occurring the event $E_{S_E}$ implies that $(\hat{s}_{1E}^{(b+1)}, \hat{s}_{2E}^{(b+1)}) = (s_1^{(b+1)}, s_2^{(b+1)})$ which means $(z_1^{(b)}, z_2^{(b)}) \in \mathcal{S}_{(s_1^{(b+1)}, s_2^{(b+1)})}$. So, the probability of error for the event $E_{z_E}$ is

$p_{E_{z_{1E}}}^e \leq$
$\mathbb{E}\left\{\sum_{(z_{1E},z_{2E}) \neq (z_1^{(b)}, z_2^{(b)})} \Pr\left((z_{1E}, z_{2E}) \in \mathcal{S}_{(s_1^{(b+1)}, s_2^{(b+1)})} | F^{(b)^c}\right)\right\} \leq$
$\mathbb{E}\{\|\mathcal{L}_E(y_0^{n(b)})\||F^{(b)^c}\}(2^{nR_{s_1}} + 2^{nR_{s_2}} + 2^{n(R_{s_1}+R_{s_2})}).$ (33)

From (32) and (33), and for sufficiently large $n$ and after the elimination of redundant inequalities, (26)-(31) are proved. ■

The receiver, to jointly decode the relays' compressed information, considers block $b-1$ and calculates his ambiguity set $\mathcal{L}_D(y_0^{n(b-1)})$ of $(\hat{z}_{1D}^{(b-1)}, \hat{z}_{2D}^{(b-1)})$ such that $(v_1^n(\hat{w}_{01D}^{(b-1)}), v_2^n(\hat{w}_{02D}^{(b-1)}), x_1^n(\hat{s}_{1D}^{(b-1)}, \hat{w}_{01D}^{(b-1)}), x_2^n(\hat{s}_{2D}^{(b-1)}, \hat{w}_{02D}^{(b-1)}), y_0^{n(b-1)}, \hat{y}_1^n(z_{1D}^{(b-1)}|\hat{s}_{1D}^{(b-1)}, \hat{w}_{01D}^{(b-1)}), \hat{y}_2^n(z_{2D}^{(b-1)}|\hat{s}_{1D}^{(b-1)}, \hat{w}_{02D}^{(b-1)}))$ are jointly $\epsilon$-typical.

The receiver declares that $(\hat{z}_{1D}^{(b-1)}, \hat{z}_{2D}^{(b-1)})$ were sent in block $b-1$ iff there is a unique pair

$(\hat{z}_{1D}^{(b-1)}, \hat{z}_{2D}^{(b-1)}) \in \mathcal{S}_{(\hat{s}_{1D}^{(b)}, \hat{s}_{2D}^{(b)})} \cap \mathcal{S}_{(\hat{w}_{012}^{(b+1)}, \hat{w}_{022}^{(b+1)})} \cap \mathcal{L}_E(y_0^{n(b-1)})$

For sufficiently large $n$, the decoding error in this step is arbitrarily small if

$\hat{R}_1 < I(\hat{Y}_1; Y_0, V_2, X_2, \hat{Y}_2|V_1, X) + I(V_1, X_1; V_2, X_2) + R_{s_1} + R_{012}$ (34)
$\hat{R}_2 < I(\hat{Y}_2; Y_0, V_1, X_1, \hat{Y}_1|V_2, X_2) + I(V_1, X_1; V_2, X_2) + R_{s_1} + R_{012}$ (35)
$\hat{R}_1 + \hat{R}_2 < I(\hat{Y}_1, \hat{Y}_2; Y_0|V_1, X_1, V_2, X_2) + I(V_1, X_1; V_2, X_2) + R_{s_1} + R_{012}$ (36)
$\hat{R}_1 < I(\hat{Y}_1; Y_0, V_2, X_2, \hat{Y}_2|V_1, X_1) + I(V_1, X_1; V_2, X_2) + R_{s_2} + R_{022}$ (37)
$\hat{R}_2 < I(\hat{Y}_2; Y_0, V_1, X_1, \hat{Y}_1|V_2, X_2) + I(V_1, X_1; V_2, X_2) + R_{s_2} + R_{022}$ (38)
$\hat{R}_1 + \hat{R}_2 < I(\hat{Y}_1, \hat{Y}_2; Y_0|V_1, X_1, V_2, X_2) + I(V_1, X_1; V_2, X_2) + R_{s_2} + R_{022}$ (39)

The proof of (34)-(39) is similar to those for (26)-(31).

The achievable rate in this case is identical to rate (1). However, the rates (26)-(31) and (34)-(39) make some relaxation on the constraints (2)-(4). thus, the achievable rate with joint decoding of the relays' compressed information can be larger than the one in theorem 1.

ACKNOWLEDGMENT

The authors wish to thank the ISSL of Sharif University of Technology and CCL of K.N. Toosi University of Technology members for their helpful comments. This work was partially supported by Iran National Science Foundation (INSF) under contract No. 88114/46-2010.